\documentclass[%
 reprint,
 superscriptaddress,
 amsmath,amssymb,
 aps,
 prl,
]{revtex4-1}

\usepackage{dsfont,amssymb,amsmath,amsthm,amsfonts,amsbsy,mathrsfs}
\usepackage{graphicx}
\usepackage{color}
\usepackage{bm}
\usepackage{multirow}
\usepackage{natbib}
\usepackage{comment}

\newcommand{\Dm}{\mathcal{ D}}
\newcommand{\pa}{\partial }
\newcommand{\LL}{{\mathbf L}}

\newcommand{\X}{\mbox{\boldmath$X$}}

\newcommand{\BEQ}{\begin{equation}}
\newcommand{\EEQ}{\end{equation}}
\newcommand{\BEA}{\begin{eqnarray}}
\newcommand{\EEA}{\end{eqnarray}}

\usepackage{tikz}
\usetikzlibrary{arrows,shapes,chains}

\begin{document}
\preprint{APS/123-QED}

\title{Exact Hamiltonian Dynamics of Rare Events in Active Matter}

\author{Andrea Crisanti}
\affiliation{Dipartimento di Fisica,
Sapienza Universit\`a di Roma, Piazzale A. Moro 2, I-00185, Rome, Italy}

\author{Matteo Paoluzzi}
\email{matteo.paoluzzi@uniroma1.it}
\affiliation{Dipartimento di Fisica,
Sapienza Universit\`a di Roma, Piazzale A. Moro 2, I-00185, Rome, Italy}
\affiliation{Istituto Nazionale di Fisica Nucleare, Sezione Roma 1, Rome, Italy}

\date{\today}

\begin{abstract}
Active systems navigate complex environments through non-equilibrium fluctuations, rendering standard equilibrium transition-rate theories inadequate. 
Moreover, transition rates provide only partial information on how stochastic dynamics explore metastable states, whereas knowledge of the optimal paths offers deeper physical insight. 
Using an active Ornstein-Uhlenbeck particle, i.e., a particle driven by exponentially correlated noise, as a paradigmatic model, 
we establish an exact mapping of the non-Markovian optimal path onto a higher-dimensional Hamiltonian dynamical system so 
that, for arbitrary force fields, optimal paths can be computed systematically by solving the corresponding Hamilton equations. 
Tuning the conserved energy allows us to explore diverse dynamical regimes, ranging from classical instanton trajectories strictly confined to the zero-energy surface to finite-time optimal paths at non-zero energies, which can spiral around shoulders of the energy landscape, a distinct signature of the non-Markovian active dynamics.
\end{abstract}

\maketitle

\paragraph*{Introduction.}
Understanding how a system navigates a complex energy landscape driven by noisy dynamics plays a fundamental role in both equilibrium and non-equilibrium statistical mechanics \cite{RevModPhys.62.251,gardiner2009stochastic,truhlar1996current}, with strong implications ranging from glassy dynamics \cite{berthier2011theoretical} and protein folding \cite{zwanzig1997two,PhysRevLett.96.228104} to astrophysics \cite{chandrasekhar1943stochastic}. 
In equilibrium statistical mechanics, 
transitions between metastable states are rare events triggered by thermal noise, leading to the well-known Arrhenius activation law. In the language of field theory, transitions between metastable states are mediated by instantons whose typical trajectories can be estimated by performing a loop expansion \cite{brezin1977perturbation}. 
Such an approach is 
twofold, as it allows us to compute both the trajectory and the transition rate. This approach can be extended to complex systems, e.g., spin glasses \cite{lopatin1999instantons,ros2021dynamical}.

However, in the biological and macroscopic world, fluctuations are predominantly non-thermal. Driven by active processes across different time and length scales, the noise in active matter is time-correlated \cite{PhysRevLett.133.118301,Maggi14}, rendering the underlying stochastic dynamics fundamentally non-Markovian and manifestly out of equilibrium \cite{PhysRevLett.117.038103}. 
Central results have been obtained in studying the escape dynamics of active particles \cite{PhysRevLett.122.258001,PhysRevLett.124.118002,caprini2019active}. However, the challenge of determining the optimal paths followed by an active system escaping a metastable state remains open, with only a few available results limited to free or harmonically confined active particles \cite{PhysRevE.107.034110,PhysRevE.107.054130,PhysRevE.106.064120}.

In this Letter, we address this question by developing a dynamical field theory for non-equilibrium stochastic dynamics applied to an Active Ornstein-Uhlenbeck Particle \cite{Szamel14,maggi2015multidimensional,MM15,Fodor16,PhysRevE.103.032607} moving in an arbitrary energy landscape. In the small-noise limit, we show that the computation of the non-Markovian most probable path exactly corresponds to a variational problem governed by a higher-dimensional Hamiltonian dynamical system. 
By embedding the dynamics into an extended Markovian phase space, we avoid the mathematical complexities of non-local actions \cite{bray1989instanton,PhysRevA.40.4050} while maintaining an exact Hamiltonian structure. 
This exact mechanical mapping allows us to unveil two striking features of active rare transitions, as well as the morphology of the optimal trajectories. 

First, we provide the computation of optimal trajectories in the small-noise limit for both finite and infinite time windows. 
The latter are the well-known instantons, i.e., trajectories moving on the zero-energy surface, but for non-Markovian stochastic dynamics. 
Moreover, we show that one can also compute optimal trajectories over a finite time window by moving on the non-zero energy surface. 
In particular, as a major novelty with respect to the Markovian case, we show that, above a threshold value of the persistence time, optimal paths can spiral around shoulders of the energy landscape. 
Additionally, we demonstrate that naive approaches based on an Arrhenius-like law and effective stationary distributions \cite{Jung87,Hanggi95} 
systematically miss the correct statistical weight of the optimal path, 
proving that effective equilibrium models fundamentally fail to capture the true non-equilibrium large deviations. 
Second, by exploring the strongly persistent limit of the active noise, we reveal a paradigm shift in the escape mechanism: the transition is no longer dictated by the saddle point of the potential energy barrier, but is instead governed by the mechanical inflection points of the energy landscape, where the deterministic force is maximized. In particular, we arrive at an exact expression that predicts the scaling of the transition rate as a function of both the height of the energy barrier and the local curvature.

\paragraph*{Non-equilibrium stochastic dynamics.}
As a model system, we consider the stochastic dynamics of a single degree of freedom $\phi \!\equiv\! \phi(t)$, described in its most general form by the following overdamped Langevin equation
\begin{align} \label{eq:stoc_model}
\dot{\phi} = \mu f(\phi) + \eta
\end{align}
where $f(\phi)$ is a deterministic drift, which we assume is produced by a potential $V(\phi)$ so that $f(\phi) \!=\! -V'(\phi)$. However, the results hold for a generic force field $f(\phi)$.
The microscopic time-scale of the relaxation dynamics is set by the inverse of the mobility $\mu\!=\!1$.
The noise term $\eta \equiv \eta(t)$ represents the effect of the environment on our variable of interest $\phi$. 
If $\eta$ is Gaussian but not delta-correlated,  
the stochastic dynamics (\ref{eq:stoc_model}) is out of equilibrium. 
For a minimal deviation from Markovian dynamics, we can take $\langle \eta(t) \rangle \!=\! 0$, and $\langle \eta(t) \eta(s) \rangle \!=\! 2 D K(|t-s|,\gamma)$. In the following, we set $K(t,\gamma) \!=\! \frac{\gamma}{2} \, e^{- \gamma |t| }$, with $\gamma$ the control parameter that tunes the non-Markovianity, i.e., the white noise limit corresponds to $\gamma \to \infty$. 
Active matter is a simple manifestation of this non-Markovian dynamics where $\gamma \!=\! \tau^{-1}$ tunes the persistence time $\tau$ of the self-propelled motion, and the self-propulsion velocity $v_{\text{self}}$ is linked to the strength of the noise $D$ through $v_{\text{self}}^2 \!=\! D / \tau$.

\begin{figure*}[!t]
\centering\includegraphics[width=.95\textwidth]{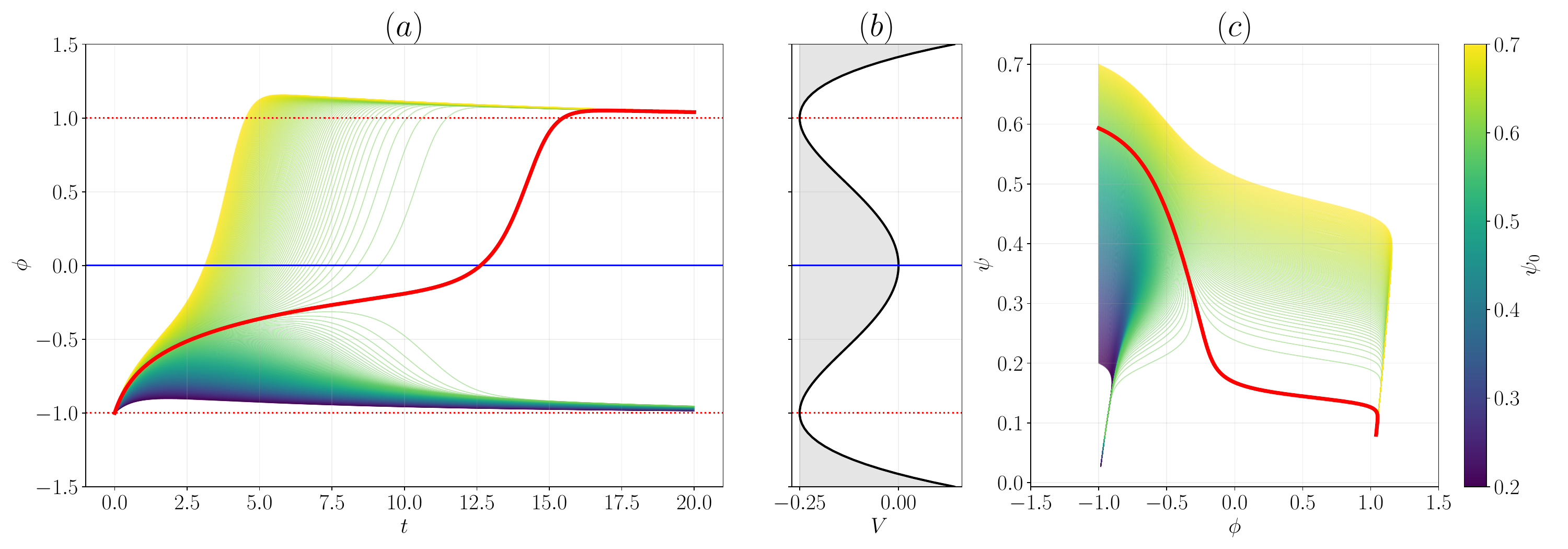}
\caption{ Zero-noise optimal trajectories of non-Markovian dynamics. (a) Optimal trajectories in the small-noise limit obtained from numerical integration of (\ref{eq:sp_eqs}). Different colors refer to different initial conditions on $\psi_0$ (i.e., different initial values of the self-propulsion force) on the energy surface $E\!=\!0$ ($\gamma=0.01$). The red curve is the closest trajectory to the boundary conditions $\phi(t_0) \!=\! \phi_0$ and $\phi(t_1) \!=\! \phi_1$. We consider a double-well potential, as depicted in (b). Panel (c) depicts the phase portrait projected onto the $(\phi,\psi)$ plane. 
}
\label{fig:1}
\end{figure*}

The final goal is to compute trajectories that maximize the transition probability $P(\phi_1,t_1 | \phi_0, t_0)$. 
To do so, we employ a dynamical field theoretic representation of the non-Markovian dynamics. 
However, instead of developing the path integral representation on top of $\eta$ (for instance, as in \cite{bray1989instanton,PhysRevA.40.4050}), we consider the extended Markovian dynamics of two degrees of freedom $(\phi,\psi)$, that is,
\begin{subequations} \label{eq:int_dyn}
\begin{align}
\dot{\phi} &= f(\phi) + \psi \\ 
\dot{\psi} &= -\gamma \psi + \xi
\end{align}
\end{subequations}
with $  \langle \xi(t) \rangle \!=\! 0$, and  $\langle \xi(t) \xi(s) \rangle \!=\! 2 D \gamma^2 \delta(t-s)$, and initial conditions $\phi_0 \!=\! \phi(t_0)$
and $\psi_0 \!=\! \psi(t_0)$. This is the standard way to represent the exponentially correlated noise.
One has $\langle \psi(t) \rangle \!=\! \psi_0 e^{-\gamma (t-t_0)} $ and $\langle \psi(t) \psi(s) \rangle \!=\! \gamma D e^{-\gamma |t-s|} + \left[ \psi_0^2 - \gamma D \right] e^{-\gamma (t + s - 2 t_0)}$
(see SI), meaning that we recover the formal expression $(\ref{eq:stoc_model})$ in the limit $\lim_{t_0 \to -\infty} \langle \psi(t) \psi(s) \rangle \!=\! \langle \eta (t) \eta (s) \rangle \!=\! 2 D K(| t -s|, \gamma)$.

\paragraph*{From stochastic to Hamiltonian dynamics.}
We can now write the transition probability from $(t_0,\phi_0)$ to $(t_1,\phi_1)$ as follows (see SI)
\begin{subequations}\label{eq:path}
\begin{align} 
     P(\phi_1,t_1 | \phi_0 ,t_0) &= \int_{\phi(t_0) =  \phi_0}^{\phi(t_1) = \phi_1} \Dm \X \, e^{-S [\X ]/D} \\ 
   S[\X] &= \int_{t_0}^{t_1} ds \, \left[\dot{\phi} p_{\phi} + \gamma^{-1} \dot{\psi} p_{\psi} - H(\X) \right] \\ 
   H(\X) &\equiv p_{\psi}^2 + p_{\phi} f(\phi) + p_{\phi} \psi - p_{\psi} \psi \; ,
\end{align}
\end{subequations}
where $\X$ indicates the set of dynamical fields over which the path integral is performed, i.e., $\X \!\equiv\! (\phi,p_{\phi},\psi,p_{\psi})$, with $\phi$ and $\psi$
the physical fields, and $p_{\phi}$ and $p_{\psi}$ the response fields that play the role of conjugate momenta. 
We see that, even though we started from a non-equilibrium stochastic dynamics, we can express the path integral in terms of a
Hamiltonian function $H(\X)$. The computation of the path integral in (\ref{eq:path}) is usually a formidable task that can be accomplished
only in the case of a quadratic action \cite{PhysRevE.107.034110}.
However, in the small noise limit $D \to 0$, the path integral
will be dominated by trajectories that minimize the dynamical action $S$ and thus maximize the transition probability, i.e., 
$ \lim_{D \to 0} P(\phi_1,t_1 | \phi_0, t_0) \!\sim\! e^{-S_{SP} / D}$ with the saddle-point action $S_{SP} \!\equiv \!S[\X_{SP}]$, where 
the trajectories $\X_{SP}$ are given by solving
\begin{align} \label{eq:SP}
 \left.\frac{\delta S}{\delta \X } \right|_{SP} = 0 \; .
\end{align}
Since they maximize the probability, such trajectories constrained by the boundary conditions
$\phi_0 \!=\! \phi(t_0)$ and $\phi_1\!=\!\phi(t_1)$ are the most probable paths. 
The equations of motion take a Hamiltonian form, with the response fields playing the role of conjugate momenta: $\dot{p_{\phi}} \!=\! - \frac{\pa H}{\pa \phi}$, $\dot{\phi} \!=\! \frac{\pa H}{\pa p_{\phi}}$,
$ \gamma^{-1}  \dot{p_{\psi}} \!=\! - \frac{\pa H}{\pa \psi}$, and $\gamma^{-1}\dot{\psi} \!=\! \frac{\pa H}{\pa p_{\psi}}$.
The explicit computation yields 
\begin{subequations} \label{eq:sp_eqs}
\begin{align} 
     \dot{p_{\phi}} &= -p_{\phi} f'(\phi) \\ 
     \dot{\phi} &= f(\phi) + \psi \\
    \gamma^{-1} \dot{ p_{\psi}} &= p_{\psi} - p_{\phi} \\ 
    \gamma^{-1} \dot{ \psi} &= 2 p_{\psi} - \psi \; .
\end{align}
\end{subequations}
We see that, in the Markovian limit, i.e.,  $\gamma \!\to\! \infty$, $\psi$ and $p_{\psi}$ relax instantaneously so that
$  p_{\psi} \!=\! p_{\phi} $, and $ p_{\psi} \!=\! \frac{\psi}{2} $. In this limit, the dynamics of the optimal path reads
$ \dot{\psi} \!=\! -\psi f'(\phi) $, and $ \dot{\phi} \!=\! f(\phi) \!+\! \psi $,
i.e., the saddle-point equations of equilibrium dynamics (see SI).

\begin{figure}[!t]
\centering\includegraphics[width=\columnwidth]{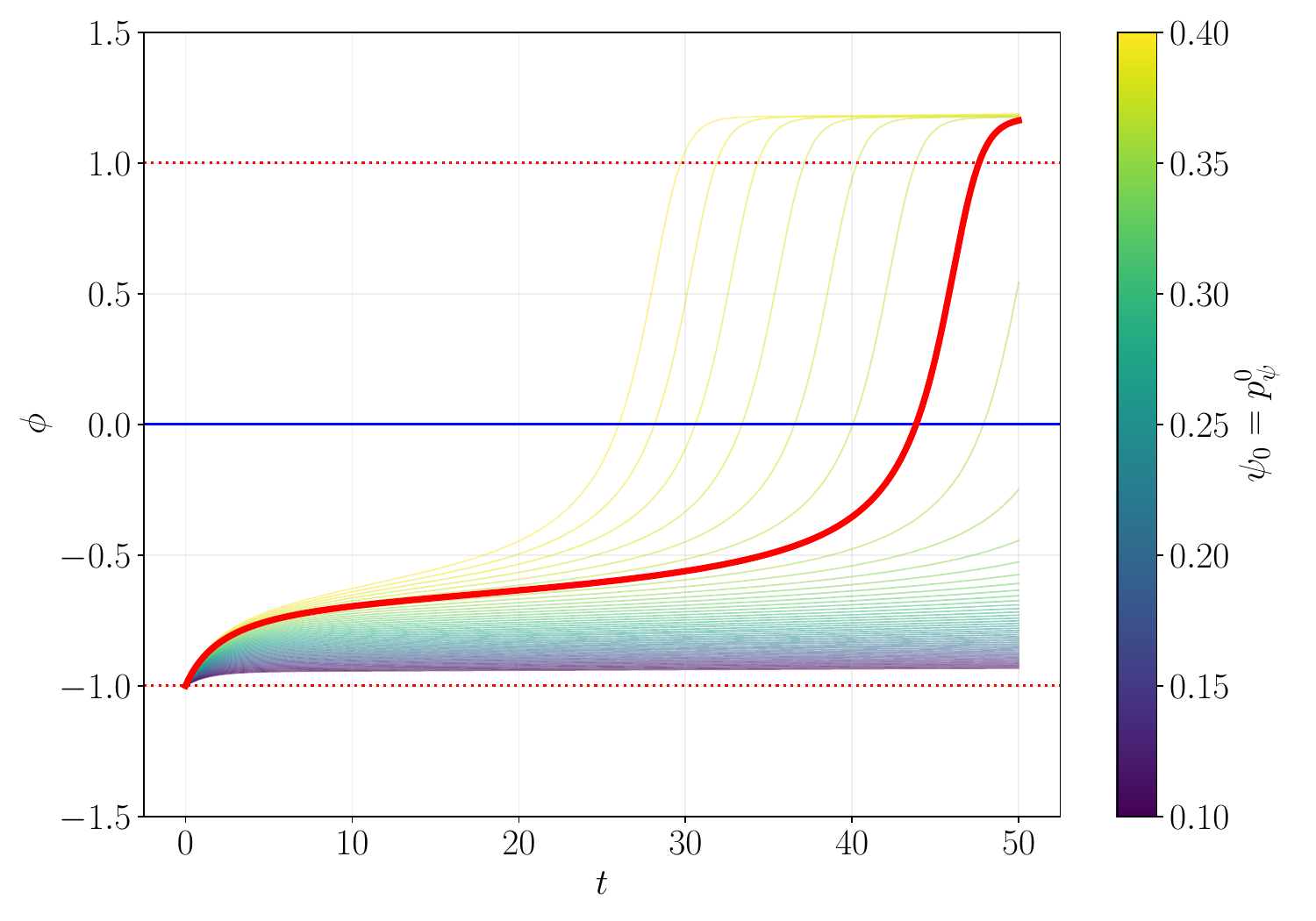}
\caption{ Optimal trajectories of non-Markovian dynamics. Trajectories $\phi(t)$ for different initial noise fluctuations $p_\psi(0) \equiv p_\psi^0$, with initial condition on the active force $\psi_0 = p_\psi^0$
so that the non-linear deterministic dynamics evolves on the zero-energy surface $E=0$ ($\gamma = 0.01$). The red curve is the trajectory capable of crossing the barrier and approaching the equilibrium point at $\phi_1=1$.
The asymptotic deviation from the target minimum reflects the extreme sensitivity of the deterministic downhill dynamics.
}
\label{fig:1b}
\end{figure}

\paragraph*{Optimal path.}  
In the Hamiltonian dynamics,
we can have orbits of period 
$T$. The instantonic trajectories correspond to orbits with $T \to \infty$ connecting mechanical critical points,
i.e., the heteroclinic orbits. Such trajectories represent rare events that take an infinite 
amount of time to take place. 
The Hamiltonian is conserved along the saddle-point dynamics, $H(\X)\!=\!E$ (see End Matter). Since $T \!=\! t_1 \!-\! t_0 \!\to\! \infty$, instantons stay on the zero-energy surface $E\!=\!p_\psi ( p_\psi \!-\! \psi) \!+\! p_\phi (f \!+\! \psi)=0$.

The solution of (\ref{eq:sp_eqs}) requires four boundary conditions. 
Using a brute force approach, i.e., the shooting problem, 
the boundary problem can be recast as follows: which initial conditions $\phi_0 \!=\! \phi(t_0)$,
$\psi(t_0)$, $p_\psi(t_0)$, and $p_\phi(t_0)$ are compatible with $\phi_1 \!=\! \phi(t_1)$? We see that $p_\psi(t_0)\!=\!p_\phi(t_0)=0$ is
a possible choice compatible with $E\!=\!0$. To this choice corresponds a purely deterministic dynamics (the generalized momenta
represent the noise). Within this set-up, the dynamics reduces to a single equation $\dot{\phi} \!=\! f(\phi) \!+\! \psi_0 e^{-\gamma (t\!-\!t_0)}$.
Starting from a local minimum, the system has to find a way to climb the energy well. In the absence of noise, this depends
on the magnitude of the self-propelling force $\psi_0$, and on the time scale $t_1 \!-\! t_0$.
If $\gamma$ is small compared with the typical relaxation rate of $f$, $\psi(t)$ becomes a quenched
variable $\psi(t) \!\simeq\! \psi_0$, while in the opposite limit $\psi(t) \!\to\! 0$; we thus have a competition between time scales. 
The escape dynamics is thus characterized by a bifurcation. We show this feature
in Fig.~(\ref{fig:1}), where we report the numerical integration of (\ref{eq:sp_eqs}) with boundary conditions  $E\!=\!p_\psi(t_0)\!=\!p_\phi(t_0)\!=\!0$.  Figure~(\ref{fig:1},a) reports
the trajectories as $\psi_0$ increases, in Fig.~(\ref{fig:1},b) we depict the potential, i.e., the standard double-well $V(\phi) \!=\! -\phi^2/2 \!+\! \phi^4 / 4$. Finally,
in Fig.~(\ref{fig:1},c), we report the phase portrait. 
On the zero energy surface we have
\begin{align} \label{eq:p_phi}
p_\phi = \frac{p_\psi ( \psi - p_\psi)}{f(\phi) + \psi}
\end{align}
and thus we can set as an initial condition $p_\psi(0) = \psi(0)$, so that the dynamics is still on the zero energy surface with $p_\phi(0)=0$
but with noise since $p_\psi(0) \neq 0$.
We report this case in Fig.~(\ref{fig:1b}), by systematically exploring the values of the initial fluctuation $p_\psi(0)$.
As in the noiseless case $p_\psi(0)\!=\!0$, we observe a sharp transition: for values below a critical threshold, the particle does not receive 
sufficient energy from the noise to reach the top of the potential, relaxing back to the initial metastable state at $\phi \!=\! -1$.
 By increasing $p_\psi(0) \!=\! \psi_0$, the non-Markovian noise allows the trajectory to cross the barrier.

We now consider a finite time window $t_1 \!-\! t_0 \!<\! \infty$, i.e.,
paths characterized by $E \!\neq\! 0$, which are generally unbounded. 
In this case, the critical points $\X_{\infty}^{(2)}$ at the inflection points (see End Matter), which do not lie on
 the zero-energy surface and are therefore relevant only for finite-time transitions, shape the dynamics according to the local form of $V(\phi)$:
they are saddle-centers where $|f|$ is maximal, and become spiral points at shoulders of the potential for $\gamma \!<\! \gamma_c$, where nearby optimal paths rotate around them.

\begin{figure}[!t]
\centering\includegraphics[width=.9\columnwidth]{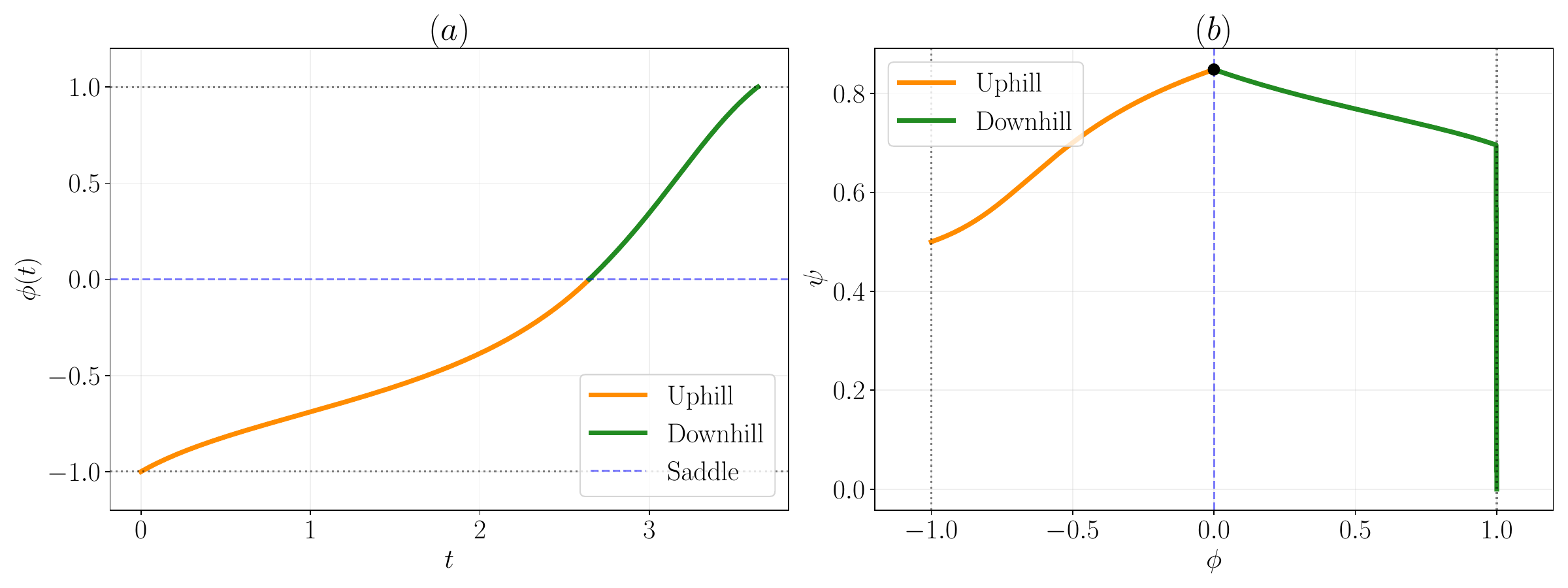}
\caption{ Instanton. (a) Uphill and downhill trajectories (\ref{eq:uphill}) and (\ref{eq:down}), respectively. (b) Active force $\psi$ as a function of $\phi$ along the two trajectories.
}
\label{fig:3}
\end{figure}

\paragraph*{Uphill \& Downhill.}
Going back to the 
solutions with $t_1 \!-\! t_0 \!\to\! \infty$ and $E\!=\!0$,
we can split the optimal path into downhill and uphill trajectories. 
Downhill paths are those of the deterministic (noiseless) dynamics, i.e., $p_\psi \!=\! p_\phi \!=\! 0$, in this case
we obtain
\begin{subequations} \label{eq:down}
\begin{align}
\dot{\phi} &= f(\phi) + \psi \\ 
\dot{\psi} &= - \gamma \psi \; .
\end{align}
\end{subequations}
The uphill trajectories are possible because of the presence of the noise, i.e., $p_\psi \!\neq\! 0$ and $p_\phi \!\neq\! 0$,
with $p_\phi$ given by (\ref{eq:p_phi}),
and thus they satisfy
\begin{subequations} \label{eq:uphill}
\begin{align}
\dot{\phi} &= f(\phi) + \psi \\ 
\gamma^{-1} \dot{p}_\psi &= \frac{p_\psi (f + p_\psi) }{f + \psi} \\
\gamma^{-1} \dot{\psi} &= 2 p_\psi - \psi \; .
\end{align}
\end{subequations}
Eliminating time, we obtain the equations for the orbits
\begin{subequations} \label{eq:orbits}
\begin{align}
\frac{d \psi}{d \phi} &= \frac{\gamma (2 p_\psi - \psi)}{f(\phi) + \psi} \\
\frac{d p_\psi}{d \phi} &= \frac{\gamma \, p_\psi (f(\phi) + p_\psi )}{(f(\phi) + \psi)^2 } \, .
\end{align}
\end{subequations}
Figure~\ref{fig:3} reports uphill/downhill trajectories and $\psi(\phi)$ obtained from (\ref{eq:orbits}). 
\paragraph*{Statistical weight of the optimal path.} 
The uphill equations provide the large deviation estimate of the leading order contribution to the statistical weight of the optimal path. 
To evaluate its cost,
we start from (\ref{eq:uphill}), which gives $dt \!=\! \frac{d\phi}{ f(\phi) \!+\! \psi}$. 
Once we plug it into the expression of $S$, we obtain
\begin{align} \label{eq:act2}
S = \int_{\phi_0}^{\phi_1} d\phi \, \frac{p_{\psi}^2 }{f(\phi) + \psi} \; .
\end{align}
Recall that the statistical weight of the optimal trajectory reads 
$\lim_{t_1 \!-\! t_0 \!\to\! \infty} P(\phi_1,t_1 | \phi_0, t_0) \!=\! P(\phi_1 | \phi_0) \!\sim\! e^{-S_{SP} / D}$. Along the downhill path, the dynamical action vanishes and thus $P(\phi_1 | \phi_0) \!\sim\! 1$.
 Along the uphill path, the weight is given by (\ref{eq:act2}). It is easy to check that, in the case of a white noise, this immediately leads to the Arrhenius law, i.e., $P_{\text{white}}(\phi_1 | \phi_0) \!\sim\! e^{-\Delta V/D }$. 
In the non-Markovian case, the probability of the uphill trajectory is 
\begin{align} \label{eq:st}
P (\phi_1 | \phi_0) \sim e^{- \frac{1}{D} \int_{\phi_0}^{\phi_1} d\phi \, \frac{p_{\psi}^2 }{f(\phi) + \psi}}
\end{align}
where $p_\psi \!=\! p_\psi (\phi)$ and $\psi \!=\! \psi(\phi)$ through (\ref{eq:orbits}). 
In the Markovian case,
the rate of decay of the probability is governed by the energy difference between the two states. There is no reason to expect a similar picture for non-equilibrium dynamics. However,
approximation schemes suggest that the stationary properties of the system might be rationalized in terms of an effective energy in some limiting situations \cite{fox1983correlation,Jung87}.
We can make contact between (\ref{eq:st}) and known approximated expressions, such as the small-$\tau$ or UCNA approximations, by performing a small-$\tau$ expansion (see SI), that yields 
$P(\phi_1 | \phi_0) \!\sim\! e^{-\Delta H_{\text{eff}} / D}$ with $H_{\text{eff}} (\phi) \!=\! V(\phi) \!+\! \tau f(\phi)^2 \!+\! O(\tau^2)$,
that is exactly what one gets from the other approximation schemes, apart from a factor of two in front of $\tau f^2$. 
This is an indication that a naive approach based on an Arrhenius-like law misses the correct statistical weight. 


\paragraph*{Small--$\gamma$ limit.} A natural question is whether an active, i.e. non-Markovian, dynamics helps in exploring a complex energy landscape. 
With this aim, we now consider the extreme case $\gamma \!\to\! 0$ ($\tau \!\to\! \infty$) in which
the system is strongly non-Markovian. 
It is worth noting that, for $\gamma\!=\!0$, the active force becomes a quenched random variable and, from the point of view of the
original stochastic dynamics, the motion is regulated by $\dot{\phi} \!=\! -V'_{\text{eff}}$,  
with $V_{\text{eff}} \!=\! V(\phi) - \psi_0 \phi$, where $\psi_0$ is the value of the active force at the initial time.
To compute the optimal path in this limit, a different rescaling of the response fields is required (see End Matter), which leads to $P(\phi_1,t_1 | \phi_0, t_0) \!\sim\! e^{-S_{\text{small}-\gamma}[\X_{SP}] / D \gamma }$, with $S_{\text{small}-\gamma} \!=\! \int dt \, \left[  p_\psi \dot{\psi} - H_{\text{small}-\gamma} \right]$ and $H_{\text{small}-\gamma} \!=\! p_\psi^2 + p_\phi (f + \psi) - p_\psi \psi$.
\begin{figure}[!t]
\centering\includegraphics[width=.9\columnwidth]{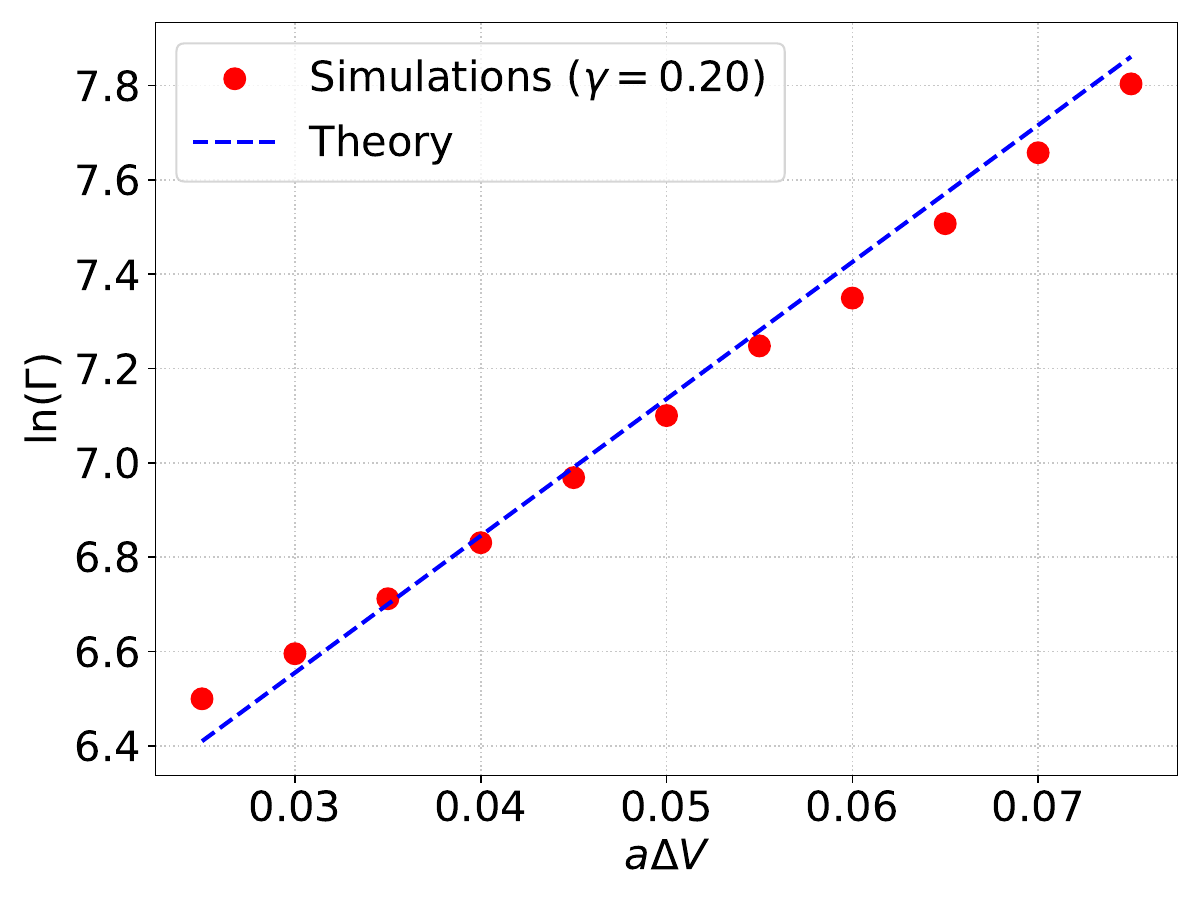}
\caption{ Transition time $\Gamma$. Test of the theoretical prediction against numerical simulations of
the non-Markovian Langevin model (red symbols), with $\Gamma \equiv A e^{ |a| \Delta V / D \gamma} $,
and $A$ a fitting parameter ($D=0.05$, $\gamma=0.2$).
}
\label{fig:5}
\end{figure}

$H_{\text{small}-\gamma}$ is conserved and the optimal paths stay on the zero-energy surface
$p_\psi (p_\psi \!-\! \psi) \!+\! p_\phi (f \!+\! \psi) \!=\! 0$ and, because of the saddle-point equation $f(\phi) \!+\! \psi \!=\! 0 $
we have $p_\psi (p_\psi \!-\! \psi) \!=\! 0 $.
The case $p_\psi \!=\!0 $ corresponds to the downhill (zero action) solution. The dynamics is noiseless and follows $\dot{\psi} \!=\! -\psi \!=\! f(\phi)$.
The uphill path corresponds to $p_\psi \!=\! \psi \!\neq\! 0$, and $\dot{\psi} \!=\! \psi \!=\! -f(\phi)$. In general we obtain $\dot{\phi}_{\text{Up/Down} } = \pm f(\phi) / f'(\phi)$
that diverges at the inflection points.
The statistical weight of the uphill path 
can be computed using 
$p_\psi = \psi$, and $dt = d\psi / p_\psi$. The uphill path starts from the bottom of the potential, i.e., $\psi_0 = 0$, and, to escape
the trap, the active force has to overcome the maximum of the force field, i.e., $\psi_1 = f( \phi^* )$, with $\phi^* = \arg\max_{\phi} |f(\phi)|$, 
that is nothing but the inflection point of the potential, $f'(\phi^*) = -V''(\phi^*) = 0$. In agreement with early studies \cite{bray1989instanton,PhysRevLett.61.7}, we obtain
$P(\phi_1 | \phi_0) \sim e^{-\psi_1^2 / 2 D \gamma} = e^{- f(\phi^* ) ^2 / 2 D \gamma}$.
We see that, differently from the white noise case where the particle needs to jump the energy barrier, in the case of a persistent noise
the transition is ruled by the inflection point that has a smaller energy cost. 
In the case of a standard double well potential $V(\phi) = a \phi^2 / 2 + b \phi^4 / 4$, with $a\!<\!0$, we get $P(\phi_1 | \phi_0) \sim  e^{-\frac{8}{27} \frac{|a| \Delta V}{\gamma D} } $, with 
$\Delta V \equiv V(0) - V(\bar{\phi})$, so that the escape time $\Gamma \sim P(\phi_1 | \phi_0)^{-1}$ grows exponentially with the energy barrier but also with the curvature ($|a|$ in this case),
as we tested against numerical simulations in Fig.~(\ref{fig:5}). 

\paragraph*{Discussion.} 
In this Letter, we established an exact mapping between the non-Markovian optimal path of an active Ornstein-Uhlenbeck particle and a higher-dimensional Hamiltonian dynamical system. 
This mapping holds for any non-Markovian dynamics driven by an exponentially correlated noise and for any type of deterministic force field $f$, even if it is not conservative. 
As in the case of Markovian dynamics, the instanton stays on the zero-energy surface. 
Exploring solutions outside the zero-energy surface means relaxing the requirement that the dynamics starts and ends in local minima, 
allowing for any path connecting the two points within a finite time window, i.e.,  any rare trajectory
that dominates on a finite timescale.
We also showed that the non-Markovian dynamics produces a richer phenomenology than its Markovian counterpart. 
In the Markovian limit the dynamics has a single degree of freedom, so that the critical points off the zero-energy surface can only be saddles or centers. 
The additional degree of freedom introduced by the active force allows for spiral critical points, 
which appear at shoulders of the potential for $\gamma<\gamma_c$, i.e., for large enough persistence time $\tau$, and have no Markovian counterpart.
 
We have computed
analytically, in the small and large $\tau$ limits, 
the transition rate between local minima. 
We demonstrated that, to first order in $\tau$, any effective-equilibrium approximation systematically miscalculates the statistical weight of rare active transitions.
These results indicate once again that, while some stationary properties of active systems can be rationalized using effective equilibrium approaches, the non-equilibrium dynamics is remarkably different even in the large time limit. 
Furthermore, by exploring the $\gamma \to 0$ limit,
we unveiled a fundamental paradigm shift in the escape mechanism. 
We proved that transition rates are no longer governed by the standard energy barriers, but rather by the inflection points of the energy landscape, leading to a profound dependence on both the barrier height and the local spatial curvature, i.e., the relaxation time grows exponentially with the local curvature of $V(\phi)$. 
It is worth noting that, because of this feature, active dynamics can be controlled to escape more efficiently from local minima if one has the possibility of tuning the topology of local basins in the complex energy landscape. These properties should be explored in future works in the context of random neural network models driven by non-Markovian noise \cite{behera2023enhanced}. 
These results also suggest a deep impact of active dynamics on the local rearrangements of active glassy systems, which certainly deserves future investigation. 
\bibliography{bib}

\section*{End Matter}
\subsection*{Energy conservation and linear stability analysis.} 
The computation of the total time derivative of $H(\X)$ yields $\frac{d H}{d t} \!=\! 0$, i.e.,  $H(t=t_0)\!=\!E$ is conserved 
during the dynamics.
Using  (\ref{eq:sp_eqs}), we have $S_{SP} \!=\! \int_{t_0}^{t_1} ds \, p_{\psi}^2(s)$, i.e.,
the energy cost of a fluctuation is proportional to the ``kinetic energy'' of the active drive.
On the other hand, using the Hamiltonian formulation, 
and considering $H(\X) \!=\! E$, we can write
\begin{align} 
S[\X] \!=\! E (t_0 - t_1) + \int_{t_0}^{t_1} ds \,  [\dot{\phi} p_{\phi} + \gamma^{-1} \dot{\psi} p_{\psi} ] \; .
\end{align}
At long times, the dynamics is dominated by its critical points obtained by setting $\dot{\X}\!=\!0$. 
There are two possibilities: $\X_{\infty}^{(1)} \!=\! (\bar{\phi},0,0,0)$, where $\bar{\phi}$ satisfies $\,f(\bar{\phi})\!=\!0$, corresponding to the mechanical critical point.
We see that this point sits on the surface $H(\X_{\infty}^{(1)}) \!=\! H (\bar{\phi},0,0,0) \!=\! 0$ and thus, since $H(\X)$ is a conserved quantity, trajectories connecting mechanical critical points 
move on such a surface.
The second critical point is $\X_{\infty}^{(2)} \!=\! (\phi^\dagger,-\frac{1}{2} f^\dagger,-f^\dagger,-\frac{1}{2}f^\dagger)$, with $\phi^\dagger: \,  f'(\phi^\dagger)\!=\!0$, i.e., the inflection point of $V(\phi)$. 
In this case, we have 
$H(\X_{\infty}^{(2)})\!=\!-\frac{1}{4}(f^\dagger)^2 \neq 0$ (unless $f^\dagger \!=\! 0$), and thus these points are not relevant for the optimal path in the limit $t_1 - t_0 \to \infty$. 
For the linear stability analysis, we set $\X \!=\! \X_{\infty}^{(1,2)} + \delta \X$. 
The perturbation $\delta \X$ evolves according to the linear system $\delta \dot{\X} \!=\! \LL \delta \X$, 
with eigenvalues (see SI for details)
$    \lambda^2 \!=\! \frac{1}{2} \left[ (f'_{\infty})^2 \!+\! \gamma^2 \pm \Delta^{1/2}\right] $, where $
    \Delta \!\equiv\! \left[ (f'_{\infty})^2 \!-\! \gamma^2 \right]^2 \!+\! 8 \gamma^2 p_{\phi}^{\infty} f''_{\infty}$, and $f'_{\infty}$, $f''_{\infty}$, $p_{\phi}^{\infty}$ are evaluated at the critical point.
For the mechanical stationary point, i.e., $f(\bar{\phi})\!=\!0$, the eigenvalues are $\lambda_{1,2} \!=\! \pm r$ and $\lambda_{3,4} \!=\! \pm \gamma$, with $r \!\equiv\! f'(\bar{\phi})$.
At the inflection points, i.e., $f'(\phi^\dagger) \!=\! 0$, one has
$ \lambda^2 \!=\! \frac{\gamma^2}{2} \left[ 1 \!\pm\! \sqrt{1 \!-\! \frac{4 f^\dagger f''^\dagger } {\gamma^2}} \right] \; .$
The nature of these points depends on the sign of $f^\dagger f''^\dagger$. If $|f|$ is maximal at the inflection point ($f^\dagger f''^\dagger \!<\! 0$), as between a minimum and a barrier, the point is a saddle-center for any $\gamma$. If instead $|f|$ has a nonzero local minimum ($f^\dagger f''^\dagger \!>\! 0$, a shoulder of the potential), a threshold $\gamma_c \!\equiv\! 2 \sqrt{f^\dagger f''^\dagger}$ appears: for $\gamma \!<\! \gamma_c$ the eigenvalues become complex and nearby optimal paths spiral around the critical point, whereas in the Markovian limit the eigenvalues are $\pm\sqrt{f^\dagger f''^\dagger}$.

\subsection*{Small-$\gamma$ limit.}
In the limit $\gamma \!\to\! 0$, a different rescaling of the response fields (see SI) leads to $\tilde{S}[\X] \!=\! \int_{t_0}^{t_1} dt \, \left[ \gamma p_\phi \dot{\phi} + p_\psi \dot{\psi} - \tilde{H} [\X] \right]$, with $\tilde{H}[\X] \!=\! p_\psi^2 + p_\phi (f + \psi) - p_\psi \psi$.
The path integral now reads $P(\phi_1,t_1 | \phi_0, t_0) \!=\! \int_{\phi_0}^{\phi_1} \Dm \X \, e^{-\tilde{S}[\X] / D \gamma} $, so that the saddle-point approximation can be performed in the limit $\gamma \!\to\! 0$, $\lim_{\gamma \to 0} P(\phi_1,t_1 | \phi_0, t_0) \!\sim\! e^{-\tilde{S}_{SP} / D \gamma }$. This leads to $S_{\text{small}-\gamma}$ and $H_{\text{small}-\gamma}$ given in the main text, plus the Hamilton equations that determine the optimal paths $\X_{SP}$ (see SI).

\bibliographystyle{rsc}

\clearpage
\newpage
\onecolumngrid
\renewcommand{\thefigure}{S\arabic{figure}}
\renewcommand{\theequation}{S\arabic{equation}}
\setcounter{figure}{0}
\setcounter{equation}{0}

\section*{Supplemental Material: Exact Hamiltonian Dynamics of Rare Events in Active Matter}

\section{Stochastic dynamics with exponentially correlated noise}
We consider the following stochastic dynamics representing, for instance, the motion of an Active Ornstein-Uhlenbeck particle in one spatial dimension
\begin{subequations} \label{eq:modello}
\begin{align}
\dot{\phi} &= f(\phi) + \psi \\ 
\dot{\psi} &= -\gamma \psi + \xi
\end{align}
\end{subequations}
with $  \langle \xi(t) \rangle = 0$, and  $\langle \xi(t) \xi(s) \rangle = 2 D \gamma^2 \delta(t-s)$. $\phi=\phi(t)$ is the relevant degree of freedom,
i.e., the position of the active particle, $\psi=\psi(t)$ represents the correlated noise over an arbitrary time scale $\tau = \gamma^{-1}$, i.e., 
the self-propulsive force. $f=f(\phi)$ represents the deterministic force that we take conservative and time-independent, i.e. $f(\phi) = -V'(\phi)$.
The dynamics is completely defined once we implement the initial conditions $\phi_0 = \phi(t_0)$, and $\psi_0=\psi(t_0)$.
Once we integrate the stochastic dynamics of $\psi$, we arrive at
\begin{subequations} \label{eq:int_dyn1}
\begin{align}
\psi(t) &= \bar{\psi}(t,t_0) + \eta(t) \\ 
\bar{\psi}(t,t_0) &\equiv \psi_0 \, e^{-\gamma (t-t_0)} \\
\eta(t) &\equiv \int_{t_0}^t ds \, e^{-\gamma (t-s)} \xi(s)
\end{align}
\end{subequations}
and thus the statistical properties of the active force read
\begin{subequations}\label{eq:int_dyn2}
\begin{align}
\langle \eta(t) \rangle &= 0 \\ 
\langle \eta(t) \eta(s) \rangle &= 2 D \Delta(t,s) \\
\Delta(t,s) &\equiv \frac{\gamma}{2} \left[ e^{-\gamma |t - s| }- e^{-\gamma (t + s - 2 t_0) } \right] \; .
\end{align}
\end{subequations}
We see that, in writing formally a non-Markovian stochastic model such as
\begin{align}
\dot{\phi} = f(\phi) + \eta
\end{align}
one implicitly considers a stationary noise, i.e., the case $t_0 \to -\infty$ in (\ref{eq:int_dyn1})
so that $\Delta(t,s)$ in (\ref{eq:int_dyn2}) equals the memory kernel $K(| t-s |, \gamma) = \frac{\gamma}{2} e^{-\gamma |t-s|}$ of the original model. 
It is also clear from (\ref{eq:int_dyn2}) that an alternative way for matching the same memory kernel 
consists in keeping $t_0$ finite, and averaging over a set of initial conditions $\psi_0 = \psi(t_0)$ taken from a Gaussian distribution  with zero mean and variance $D \gamma$.

\section{Path integral}
We consider the stochastic dynamics of a single degree of freedom $\phi(t)$, with $\phi \in \mathbb{R}$, and $t \in [t_0,t_1]$, with boundary
conditions $\phi(t_0) = \phi_0$, and $\phi(t_1) = \phi_1$. In particular, we want to compute the most probable trajectories connecting these two points in the presence of a 
non-Markovian noise $\psi$ that undergoes
an Ornstein-Uhlenbeck process.
The most probable trajectories are those that maximize the transition probability $P(\phi_1,t_1 | \phi_0 ,t_0)$.
To compute the transition probability $P(\phi_1,t_1 | \phi_0 ,t_0)$, we develop a path integral formalism \cite{PhysRevA.8.423,dominicis1976technics,janssen1976lagrangean} for the following stochastic dynamics 
\begin{align}
\dot{\phi} &= f(\phi) + \psi \\ 
\dot{\psi} &= -\gamma \psi + \xi
\end{align}
with $\psi$ being the active force, and $\xi$ a white noise such that 
\begin{align}
    \langle \xi(t) \rangle = 0 \;\; , \;\; \langle \xi(t) \xi(s) \rangle = 2 D \gamma^2 \delta(t-s) \; .
\end{align}
Using Ito prescription, we write the path integral as follows
\begin{align}
P(\phi_1, t_1 | \phi_0 , t_0) = \int_{\phi_0 = \phi(t_0)}^{\phi_1 = \phi(t_1)} \Dm \xi \Dm \psi  \Dm \phi \, e^{-\frac{1}{2} \int ds \, \frac{\xi^2}{2 D \gamma^2} } \delta \left[ \dot{\phi} - f(\phi) - \psi \right] \delta \left[ \dot{\psi} + \gamma \psi - \xi \right]
\end{align}
once we introduce the response fields $p_\phi$, and $p_\psi$, we get
\begin{align}
P(\phi_1, t_1 | \phi_0 , t_0) = \int_{\phi_0 = \phi(t_0)}^{\phi_1 = \phi(t_1)} \Dm \xi \Dm \psi  \Dm p_{\psi} \Dm \phi \Dm p_{\phi} \, e^{ \int ds \, \left[-\frac{\xi^2}{4 D \gamma^2} -p_{\phi} (\dot{\phi} - f(\phi) - \psi) -p_{\psi} (\dot{\psi} + \gamma \psi - \xi) \right] } \; .
\end{align}
We thus perform the Gaussian functional integral over the noise $\xi$ arriving at the following Martin-Siggia-Rose dynamical action
(we recall that convergence of the integral is guaranteed by the fact that the integral is performed along $i p_\psi$ and $i p_\phi$)
\begin{align}
S[\phi,p_{\phi},\psi,p_{\psi}] = \int ds \, \left[ -D \gamma^2 p_{\psi}^2 + p_{\phi} (\dot{\phi} - f-\psi) + p_{\psi}(\dot{\psi} + \gamma \psi) \right] \; .
\end{align}
Once we perform the rescaling $p_{\psi} \to p_{\psi} /D\gamma$, and $p_{\phi} \to p_{\phi} / D$, we arrive at
\begin{align}
    P(\phi_1,t_1 | \phi_0 ,t_0) &= \int \Dm [\phi,p_{\phi},\psi,p_{\psi}] \, e^{-S/D} \\ 
    S[\phi,p_{\phi},\psi,p_{\psi}] &\equiv \int_{t_0}^{t_1} ds \, \left[ -p_{\psi}^2 + p_{\phi}(\dot{\phi} -f - \psi) + p_{\psi} (\gamma^{-1} \dot{\psi} + \psi )\right] \; .
\end{align}
\subsection{Markovian limit}
The white noise limit is recovered for $\gamma \to \infty$ and thus $\gamma^{-1} \to 0$. We see that
\begin{align}
    \lim_{\gamma \to \infty } S[\phi,p_{\phi},\psi,p_{\psi}] = \int_{t_0}^{t_1} ds \, \left[ -p_{\psi}^2 + p_{\phi}(\dot{\phi} -f - \psi) + p_{\psi}  \psi \right] \; .
\end{align}
We can thus integrate over $p_{\psi}$, arriving at
\begin{align}
    A[\phi,p_{\phi},\psi] = \int_{t_0}^{t_1} ds \, \left[ \frac{1}{4}\psi^2 + p_{\phi}(\dot{\phi} -f - \psi) \right] \; ,
\end{align}
and thus $\psi$ acts as a white noise on $\phi$. In particular, once we integrate over $\psi$ we arrive at
\begin{align}
P(\phi_1,t_1|\phi_0,t_0) &= \int_{\phi_0}^{\phi_1} \Dm[\phi,p_{\phi}] \, e^{-A/D} \\ 
A[\phi,p_{\phi}] &\equiv \int_{t_0}^{t_1} ds \, \left[ -p_{\phi}^2 + p_{\phi} (\dot{\phi} - f) \right] \; .
\end{align}
We see that $A[\phi,p_{\phi}]$ is the dynamical action that we obtain for a Langevin equation $\dot{\phi} = f + \eta$, with $\langle \eta(t) \rangle = 0$,
and $\langle \eta(t) \eta(s) \rangle = 2 D \delta(t-s)$, i.e., the white noise limit of the original non-Markovian Langevin model.

\section{Small noise limit}
We now consider the computation of the path that maximizes the transition probability in the small noise limit, i.e., $D \to 0$, so that the transition probability reads
\begin{align}
    P(\phi_1,t_1 | \phi_0,t_0) \sim e^{-S_{SP} / D}
\end{align}
with $S_{SP}$ the value of the dynamical action obtained along the saddle-point trajectories. 
In this limit, we thus arrive at the optimal path that is obtained by solving the following set of saddle-point equations
\begin{align}
    \left.\frac{\delta S}{\delta \phi }\right|_{\text{SP}} =  \left.\frac{\delta S}{\delta p_{\phi} }\right|_{\text{SP}} = \left.\frac{\delta S}{\delta \psi}\right|_{\text{SP}} = \left.\frac{\delta S}{\delta p_{\psi}}\right|_{\text{SP}} =0 \; .
\end{align}
Before writing the equations, we see that the dynamical action can be written in the following form
\begin{align}
    S[\phi,p_{\phi},\psi,p_{\psi}] &= \int_{t_0}^{t_1} ds \, \left[\dot{\phi} p_{\phi} + \gamma^{-1} \dot{\psi} p_{\psi} - H\right] \\ 
    H[\phi,p_{\phi},\psi,p_{\psi}] &\equiv p_{\psi}^2 + p_{\phi} f(\phi) + p_{\phi} \psi - p_{\psi} \psi \; .
\end{align}
The ``classical'' equations of motion are given by
\begin{align}
\frac{\delta S}{\delta \phi} &= -\pa_tp_{\phi} - \frac{\pa H}{\pa \phi}= 0 \\
\frac{\delta S}{\delta p_{\phi}} &= \pa_t \phi - \frac{\pa H}{\pa p_{\phi}}= 0 \\
\frac{\delta S}{\delta \psi} &= -\gamma^{-1} \pa_t p_{\psi}- \frac{\pa H}{\pa \psi}= 0 \\
\frac{\delta S}{\delta p_{\psi}} &= \gamma^{-1}\pa_t \psi - \frac{\pa H}{\pa p_{\psi}}= 0 
\end{align}
and thus we have
\begin{align}
     \dot{p}_{\phi} &= -p_{\phi} f'(\phi) \\ 
     \dot{\phi} &= f(\phi) + \psi \\
    \gamma^{-1}  \dot{p}_{\psi} &= p_{\psi} - p_{\phi} \\ 
    \gamma^{-1}  \dot{\psi} &= 2 p_{\psi} - \psi \; .
\end{align}
We see that, in the Markovian limit where $\gamma \to \infty$, $\psi$ and $p_{\psi}$ relax instantaneously so that
\begin{align}
    p_{\psi} &= p_{\phi} \\ 
    p_{\psi} &= \frac{\psi}{2}
\end{align}
and thus the dynamics of the optimal path reduces to
\begin{align}
    \dot{\psi} &= -\psi f'(\phi) \\ 
    \dot{\phi} &= f(\phi) + \psi
\end{align}
that are nothing but the saddle-point equations we get in the case of a white noise.
We now compute the total time derivative of $H$ that is given by
\begin{align}
    \frac{d H}{dt} &= \frac{\pa H}{\pa \psi}\dot{\psi} + \frac{\pa H}{\pa p_{\psi}}\dot{p_{\psi}}+ \frac{\pa H}{\pa \phi}\dot{\phi} + \frac{\pa H}{\pa p_{\phi}}\dot{p_{\phi}} =  \\ \nonumber 
&=    \gamma \frac{\pa H}{\pa \psi}\frac{\pa H}{\pa p_{\psi}} -\gamma \frac{\pa H}{\pa p_{\psi}}\frac{\pa H}{\pa \psi}+ \frac{\pa H}{\pa \phi}\frac{\pa H}{\pa p_{\phi}} - \frac{\pa H}{\pa p_{\phi}}\frac{\pa H}{\pa \phi} = 0 \; .
\end{align}
We see that energy is conserved so that, for a finite time interval, we have $H = \text{const.} = E$ and thus the dynamical action is
\begin{align} \label{eq:dynS_1}
    S = (t_0 - t_1) E + \int_{t_0}^{t_1} ds \, (\dot{\phi} p_{\phi} + \gamma^{-1} \dot{\psi} p_{\psi}) \; .
\end{align}
Moreover, we notice that, using Hamilton equations, we also have the following expression for the dynamical action
\begin{align} \label{eq:dynS_2}
    S = \int_{t_0}^{t_1} ds \, p_{\psi}(s)^2 \; .
\end{align}
To make progress, we study the critical points of the dynamics and their linear stability. The critical points $(\phi^*,p_{\phi}^*,\psi^*,p_{\psi}^*)$ satisfy the set of equations
\begin{align}
    -p_{\phi}^* f'(\phi^*) &= 0 \\ 
    f(\phi^*) + \psi^* &= 0 \\
    p_{\psi}^* - p_{\phi}^* &= 0 \\ 
    2 p_{\psi}^* - \psi^* &= 0 
\end{align}
and thus we have two possibilities $(\bar{\phi},0,0,0)$, with $\bar{\phi}:\,f(\bar{\phi})=0$, and $(\phi^\dagger,-\frac{1}{2} f^\dagger,-f^\dagger,-\frac{1}{2}f^\dagger)$, with $\phi^\dagger: \, f'(\phi^\dagger)=0$, i.e., the inflection point of the potential $V(\phi)$. We also see that for $\bar{\phi}$ we have
$H=0$, while for $\phi^\dagger$ one has $H=-\frac{1}{4}(f^\dagger)^2 \neq 0$ (unless $f^\dagger = 0$).
Next, we discuss the linear stability of these stationary points. To do so, we arrange the four fields into a vector $\X \equiv (\phi,p_{\phi},\psi,p_{\psi})$ and look at the fate of a perturbation $\delta \X \equiv \X(t) - \X^*$ under the linearized dynamics, with $\X^*$ being the critical point under consideration. The evolution of $\delta \X$ reads
\begin{align}
    \delta \dot{\X} &= \LL \delta \, \X \\
\LL &= \begin{bmatrix}
        f'(\phi^*) & 0 & 1 & 0 \\
        -p_{\phi}^* f''(\phi^*) & -f'(\phi^*) & 0 & 0 \\
        0 & 0 & -\gamma & 2\gamma \\
        0 & -\gamma & 0 & \gamma
          \end{bmatrix}
\end{align}
with eigenvalues 
\begin{align}
    \lambda^2 &= \frac{1}{2} \left[ (f')^2 + \gamma^2 \pm \Delta^{1/2}\right] \\
    \Delta &\equiv \left[ (f')^2 - \gamma^2 \right]^2 + 8 \gamma^2 p_{\phi}^* f'' \; ,
\end{align}
where $f'$ and $f''$ are evaluated at $\phi^*$.
For the mechanical stationary point characterized by $f(\bar{\phi})=0$, defining $r \equiv f'(\bar{\phi})$, we have
\begin{align}
\lambda^2 = \frac{1}{2} \left[ r^2 + \gamma^2 \pm (r^2 - \gamma^2) \right]
\end{align}
and thus
\begin{align}
\lambda_{1,2} &= \pm r \\
\lambda_{3,4} &= \pm \gamma \; .
\end{align}
For the inflection points, characterized by $f'(\phi^\dagger) = 0$, the critical point is $\X=(\phi^\dagger, -\frac{1}{2} f^\dagger, -f^\dagger, -\frac{1}{2} f^\dagger)$, and thus, once we define $f^\dagger \equiv f(\phi^\dagger)$ and $f^{\prime\prime\dagger} \equiv f''(\phi^\dagger)$, the term $\Delta$ simplifies to
\begin{align}
    \Delta = \gamma^4 - 4 \gamma^2 f^\dagger f^{\prime\prime\dagger} \; .
\end{align}
Consequently, the eigenvalues are determined by
\begin{align}
    \lambda^2 = \frac{\gamma^2}{2} \left[ 1 \pm \sqrt{1 - \frac{4 f^\dagger f^{\prime\prime\dagger}}{\gamma^2}} \right] \; .
\end{align}
The nature of this critical point depends on the sign of $f^\dagger f^{\prime\prime\dagger}$, as discussed in the End Matter of the main text.


\section{Downhill and Uphill solution}
In this section we compute the uphill and downhill trajectories obtained by considering the solutions of the saddle point equations
on the zero-energy surface $E = 0$. To do so, we start from the equations of motion
\begin{align}
\dot{p}_\phi &= -p_\phi f'(\phi)\\ 
\dot{\phi} &= f(\phi) + \psi \\
\gamma^{-1}\dot{p}_\psi &= p_\psi - p_\phi \\
\gamma^{-1} \dot{\psi} &=2 p_\psi - \psi 
\end{align}
along the energy surface $E=0$. We remind that the dynamical action and the Hamiltonian are
\begin{align}
S &= \int_{t_0}^{t_1} ds \, \left[ \dot{\phi} p_\phi + \gamma^{-1} \dot{\psi} p_\psi - H \right] = \int_{t_0}^{t_1} ds \, p_{\psi}^2 \\ 
H &= p_{\psi}^2 + p_\phi f(\phi) + p_\phi \psi - p_\psi \psi \; .
\end{align}
Moreover, the computation of the instanton implies $t_0 \to -\infty$, and $t_1 \to \infty$, with $\phi_0 = \phi(-\infty)$ and $\phi_1 = \phi(\infty)$ stationary points of the dynamics.
Since $E=0$, we also have
\begin{align}
S = \int_{t_0}^{t_1} ds \, \left[ \dot{\phi} p_\phi + \gamma^{-1} \dot{\psi} p_\psi \right] \; .
\end{align}
We notice that the condition $E=0$ implies
\begin{align}
p_\psi(p_\psi - \psi) + p_\phi (f + \psi) = 0 \; .
\end{align}
We start with the zero-noise trajectories characterized by $p_\psi = p_\phi = 0$, and thus we arrive at the downhill path
\begin{align}
\dot{\phi} &= f(\phi) + \psi \\ 
\dot{\psi} &= - \gamma \psi 
\end{align}
that is nothing but the deterministic dynamics obtained by setting to zero the non-thermal noise in the original Langevin model.
More in general, we notice that, because of $E=0$ one has
\begin{align}
p_\phi = \frac{p_\psi (\psi - p_\psi) }{f(\phi) + \psi }
\end{align}
once we plug this expression (considering the noisy case $p_\psi \neq 0$), we arrive at
the equations of motion for the uphill trajectories
\begin{align}
\dot{\phi} &= f(\phi) + \psi \\
\gamma^{-1} \dot{p}_\psi &= \frac{p_\psi (f + p_\psi) }{f + \psi} \\
\gamma^{-1} \dot{\psi} &= 2 p_\psi - \psi \; .
\end{align}
Eliminating time, we obtain the equations for the orbits
\begin{align}
\frac{d \psi}{d \phi} &= \frac{\gamma (2 p_\psi - \psi)}{f(\phi) + \psi} \\
\frac{d p_\psi}{d \phi} &= \frac{\gamma \, p_\psi (f(\phi) + p_\psi )}{(f(\phi) + \psi)^2 } \, .
\end{align}
In other words, one has
\begin{align}
dt = \frac{d\phi}{ f(\phi) + \psi}
\end{align}
that we can plug into the expression of the dynamical action obtaining
\begin{align} \label{eq:act}
S = \int_{\phi_0}^{\phi_1} d\phi \, \frac{p_{\psi}^2 }{f(\phi) + \psi} \; .
\end{align}
We remind that $S$ provides the statistical weight of the path. Considering uphill trajectories, this means 
that we have access to the stationary distribution in the small noise limit that is
\begin{align}
P_{st} (\phi) \sim e^{-\frac{1}{D} \int^{\phi} \, d\phi \, \frac{p_{\psi}^2 }{f(\phi) + \psi} }
\end{align}
To determine the prefactor, one has to go beyond the saddle point, e.g., within a one-loop approximation. 
Before doing that, we consider a large $\gamma$ limit, i.e., small $\tau = \gamma^{-1}$, to make contact between this expression and other approximation schemes that allow for the computation of the stationary distribution of models with correlated noise.
We thus write
\begin{align}
p_\psi &= p_0 + \tau p_1 + O(\tau^2) \\ 
\psi &= \psi_0 + \tau \psi_1 
\end{align}
Using the equations 
\begin{align}
\tau \dot{p}_\psi &= \frac{p_\psi (f + p_\psi) }{f + \psi} \\
\tau \dot{\psi} &= 2 p_\psi - \psi \; .
\end{align}
the zero order produces
\begin{align}
2 p_0 - \psi_0 &= 0 \\
\frac{\psi_0}{2} (f + \frac{\psi_0}{2} ) &= 0 \; ,
\end{align}
and thus we have
\begin{align} \label{eq:zero}
p_0 &= \frac{\psi_0}{2} \\ 
\psi_0 &= -2 f \; .
\end{align}
We see that, if we plug (\ref{eq:zero}) into (\ref{eq:act}), we obtain
\begin{align}
S = \int_{\phi_0}^{\phi_1} d\phi \, (-f(\phi)) + O(\tau) = V(\phi_1) - V(\phi_0)
\end{align}
and thus at the zero order we have just the Boltzmann distribution
\begin{align}
P_{st}(\phi) \sim e^{-V(\phi) / D} \; .
\end{align}
We now move to the first order, to do so we remind that from (\ref{eq:zero}) one has
\begin{align}
\psi_0 &= -2 f \\ 
p_0 &= -f \; .
\end{align}
We start from
\begin{align}
\tau \frac{d\psi}{d \phi} (f + \psi) = 2 p_\psi - \psi
\end{align}
from the left hand side we have
\begin{align}
\tau (-2 f') (f - 2 f) = 2 \tau f f'
\end{align}
while from the right hand side one has
\begin{align}
2 (p_0 + \tau p_1) - (\psi_0 + \tau \psi_1)
\end{align}
and we finally get
\begin{align}
2 p_1 - \psi_1 = 2 f f' \; .
\end{align}
From the other equation we have
\begin{align}
\tau \frac{d p_\psi}{d \phi} (f + \psi)^2 = p_\psi (f + p_\psi)
\end{align}
and thus
\begin{align}
-\tau f' f^2 = (p_0 + \tau p_1) (f + p_0 + \tau p_1)
\end{align}
so that
\begin{align}
p_1 &= f' f \\ 
\psi_1 &= 0 \; .
\end{align}
Up to the first order, we finally have
\begin{align}
\psi &= \psi_0 + O(\tau^2) = -2 f + O(\tau^2) \\
p_\psi &= p_0 + \tau p_1  + O(\tau^2) = -f + \tau f' f + O(\tau^2) \; .
\end{align}
Once we plug into the dynamical action we get
\begin{align}
S \sim \int_{\phi_0}^{\phi_1} d\phi \, \left[  -f(\phi) + 2 \tau f(\phi) f'(\phi) \right] + O(\tau^2) \; .
\end{align}
but $2 f f' = \frac{d}{d \phi} f^2$ and thus we arrive at
\begin{align}
S \simeq V(\phi_1) - V(\phi_0) + \tau \Delta f^2 + O(\tau^2)
\end{align}
with $\Delta f^2 \equiv f(\phi_1)^2 - f(\phi_0)^2$. For small $D$ and small $\tau$, this should be a better approximation than an Arrhenius-like expression within the Fox or UCNA schemes.

\section{Large persistence time limit}
We start from the dynamical action written as
\begin{align} \label{eq:pre_res}
S = \int ds \, \left[ -D \gamma^2 p_\psi^2 + p_\phi (\dot{\phi} - f - \psi) + p_\psi (\dot{\psi} + \gamma \psi) \right] \; .
\end{align}
To cleanly discuss the limit $\gamma \to 0$, we want to write
\begin{align}
P(\phi_1 | \phi_0) \sim e^{-S / D \gamma}
\end{align}
to do so, we go back to (\ref{eq:pre_res}) and we first rescale the time with $\gamma$, i.e., $t = \gamma s$, obtaining
\begin{align}
dt &= \gamma ds \\ 
\frac{d}{ds} &= \gamma \frac{d}{dt}
\end{align}
once we do so, we arrive at
\begin{align}
S = \int_{t_0}^{t_1} \frac{dt}{\gamma} \left[ -D \gamma^2 p_\psi^2 + p_\phi(\gamma \dot{\phi} -f - \psi) + p_\psi (\gamma \dot{\psi} + \gamma \psi ) \right] \; .
\end{align}
Now we perform the following rescaling $p_\psi \to p_\psi / \gamma D$, and $p_\phi \to p_\phi / D$, and we finally arrive at
\begin{align}
S &= \int dt \, \left[ \gamma p_\phi \dot{\phi} + p_\psi \dot{\psi} - H \right] \\ 
H &= p_\psi^2 + p_\phi (f + \psi) - p_\psi \psi \, .
\end{align}
We can now consider the limit $\gamma \to 0$, which leads to
\begin{align}
P(\phi_1 | \phi_0) &\sim e^{-S_{SP} / D \gamma} \\ 
S &= \int dt \, \left[  p_\psi \dot{\psi} - H \right] \\ 
H &= p_\psi^2 + p_\phi (f + \psi) - p_\psi \psi \, .
\end{align}
with the saddle point equations
\begin{align}
p_\phi f'(\phi) &= 0 \\
f(\phi) + \psi &= 0 \\ 
\dot{p}_\psi &= p_\psi - p_\phi \\
\dot{\psi} &= 2 p_\psi - \psi \; . 
\end{align}
We see that the dynamical action at the saddle point reads
\begin{align}
S = \int dt \, p_\psi^2 \; .
\end{align}
Looking at the zero energy trajectories one has
\begin{align}
p_\psi (p_\psi - \psi) + p_\phi (f + \psi) = 0
\end{align}
because of the saddle point equation $f + \psi = 0$ we have
\begin{align}
p_\psi (p_\psi - \psi) = 0
\end{align}
and thus we have two possibilities $p_\psi =0 $ that is a downhill solution since the action is zero and the dynamics is
\begin{align}
\dot{\psi} = -\psi = f(\phi)
\end{align}
and $p_\psi = \psi \neq 0$, the uphill solution 
\begin{align}
\dot{\psi} = \psi = -f(\phi) \; .
\end{align}
The statistical weight of the optimal uphill path in the $\gamma \to 0$ limit can be computed using the fact that
$p_\psi = \psi$, and $dt = d\psi / p_\psi$. The uphill path starts from the bottom of the potential, and thus we have $\psi_0 = 0$, and, to escape
the trap, the active force has to overcome the maximum of the force field between the minimum and the barrier top, i.e., $\psi_1 = f(\phi^*)$, with $\phi^* = \arg\max_{\phi} |f(\phi)|$ (that is nothing but the inflection point of the potential). We thus finally arrive at
\begin{align}
P(\phi_1 | \phi_0) \sim e^{-\psi_1^2 / 2 D \gamma} = e^{-f(\phi^*)^2 / 2 D \gamma} \; .
\end{align}

\section{Numerical implementation} \label{app:numerics_ode}
The saddle-point equations have been solved on a finite time window $t \in [t_0, t_1]$, by setting $t_0=0$, using a shooting method and thus
by an explicit integration of the set of non-linear differential equations with initial condition on the physical degree of freedom $\phi_0 = \phi(t_0)$ and
then by changing the initial conditions on $\psi_0$, $p_\psi^0$, and $p_\phi^0$. In particular, we specialized our computation to the case $E=0$ with 
$p_\psi^0 = p_\phi^0=0$, i.e., the noiseless case, and to the case $E=0$, $p_\phi^0=0$, and $\psi_0 = p_\psi^0$.
Once we mesh the time on a grid, we arrive at the following equations.
For $n=2,3,...$, we have
\begin{align*}
    p_{\phi}^n &= p_{\phi}^{n-1} e^{-\frac{dt}{2} [3 f^\prime_{n-1} - f^\prime_{n-2}]} \\
    p_{\psi}^n &= e^{\gamma dt} p_{\psi}^{n-1} - \gamma \frac{dt}{2} \left[ 3 e^{\gamma dt} p_{\phi}^{n-1} - e^{2 \gamma dt} p_{\phi}^{n-2} \right]\\
    \psi_n &= e^{-\gamma dt} \psi_{n-1} + 2 \sinh(\gamma dt) p_{\psi}^{n-1} \\ \nonumber 
    &- \gamma dt \left[ 3 \sinh(\gamma dt) p_{\phi}^{n-1} -  \sinh(2 \gamma dt) p_{\phi}^{n-2}\right] \\
    \phi_n &= \phi_{n-1} + \frac{1}{\gamma} (1-e^{-\gamma dt})\psi_{n-1} + \frac{2}{\gamma}(\cosh(\gamma dt) -1)p_{\psi}^{n-1} \\ \nonumber
    &+\frac{dt}{2} [3 f_{n-1} - f_{n-2}] \\ \nonumber 
    &- dt [ 3 (\cosh(\gamma dt) - 1) p_{\phi}^{n-1} - (\cosh(2 \gamma dt) - 1) p_{\phi}^{n-2}]
\end{align*}
while for $n=1,$ one has
\begin{align*}
    p_{\phi}^1 &= e^{-dt f^\prime_0} p_{\phi}^0\\
    p_{\psi}^1 &= e^{\gamma dt} p_{\psi}^0 -\gamma \frac{dt}{2} \left[e^{\gamma dt} p_{\phi}^0 + p_{\phi}^1 \right]\\
    \psi_1 &= e^{-\gamma dt} \psi_0 + 2 \sinh (\gamma dt) p_{\psi}^0 - \gamma dt \sinh(\gamma dt) p_{\phi}^0 \\
    \phi_1 &= \phi_0 + \frac{1}{\gamma} (1 - e^{-\gamma dt}) \psi_0 + \frac{2}{\gamma}(\cosh(\gamma dt) - 1) p_{\psi}^0 + dt f_0  \\ \nonumber 
    &- dt (\cosh(\gamma dt) - 1) p_{\phi}^0
\end{align*}
With the boundary conditions
\begin{align*}
    \phi_0 &= \phi(t_0) \\
    p_{\psi}^0 &= 0 \;\; \text{or} \;\; p_{\psi}^0 = \psi_0 \\
    p_{\phi}^0 &= \frac{1}{f_0 + \psi_0} \left[ E + p_{\psi}^0 \psi_0 - (p_{\psi}^0)^2\right]
\end{align*}
and
\begin{align*}
    f^\prime_n &\equiv f^\prime(\phi_n) \\
    f(\phi) &= -V^\prime(\phi) \\
\end{align*}
In the case of a double well potential we set $V(\phi) = \frac{a}{2} \phi^2 + \frac{b}{4} \phi^4$, with $a=-1$, and $b=1$.
In this picture, we vary the initial condition on $\psi_0$, i.e., the initial velocity, and look at the resulting trajectory $\phi(t)$. 
The final state $\phi(t_1) = \phi_1$ depends on the choice we made on $\psi_0$: by moving $\psi_0$ within an ensemble of
initial conditions taken from a uniform distribution, we look at the trajectory that satisfies the fixed boundary problem.

\section{Numerical simulations}
To compare the saddle-point dynamics with the actual stochastic dynamics, we numerically integrated Eq.~(\ref{eq:modello}) using the Euler-Maruyama method. The discretized dynamics reads
\begin{subequations}
\begin{align}
\phi_{n+1} &= \phi_n + (f_n + \psi_n) \Delta t \\
\psi_{n+1} &= \psi_n - \gamma \psi_n \Delta t + \gamma \sqrt{2 D \Delta t} \, r_n
\end{align}
\end{subequations}
where $\Delta t=10^{-2}$ is the integration time step, $f_n \equiv f(\phi_n)$, and $r_n$ is a normally distributed random number with zero mean and unit variance. To test the escape dynamics in the small-$\gamma$ limit, we performed numerical simulations of $N = 5 \times 10^{3}$ independent walkers starting at $\phi_0 = -1$ and $\psi_0 = 0$. The remaining parameters were set to $D = 0.05$ and $\gamma=0.2$.

\end{document}